\documentclass[letterpaper,pra,twocolumn,showpacs,superscriptaddress]{revtex4}

\usepackage{graphicx}
\usepackage{amsmath}
\usepackage{amssymb}

\begin{document}

\title{Continuous Variable Entanglement and Squeezing of Orbital Angular Momentum States}

\author{M. Lassen$^{1,2}$}
\noaffiliation{}

\author{G. Leuchs$^{2,3}$}
\noaffiliation{}

\author{U. L. Andersen}
\noaffiliation{}

\email{mlassen@fysik.dtu.dk}

\affiliation{Department of Physics, Technical University of Denmark, Fysikvej, 2800 Kgs. Lyngby, Denmark\\
$^{2}$Max Planck Institute for the Science of Light, G\"unther Scharowskystrasse 1, 91058 Erlangen, Germany\\
$^{3}$University Erlangen-N\"urnberg, Staudtstrasse 7/B2, 91058 Erlangen, Germany.}

\date{\today}

\begin{abstract}
We report the first experimental characterization of the first-order continuous variable orbital angular momentum states. Using a spatially non-degenerate optical parametric oscillator (OPO) we produce quadrature entanglement between the two first-order Laguerre-Gauss modes. The family of OAM modes is mapped on an orbital Poincare sphere, where the modes position on
the sphere is spanned by the three orbital parameters. Using a
non-degenerate OPO we produce squeezing of these parameters, and
as an illustration, we reconstruct the "cigar-shaped" uncertainty
volume on the orbital Poincare sphere.
\end{abstract}


\maketitle

Propagating light beams have a spin angular momentum and an orbital angular momentum (OAM), which are associated with the
polarization and the phase distribution of the light state, respectively~\cite{Allen1992}. The {\it quantum} spin angular
momentum (or polarization) has been thoroughly investigated both for discrete and for continuous variables (CV) ~\cite{Chirkin1993,Korolkova2002,Bowen2002}. Recently the {\it quantum} OAM of light has attracted a lot of attention, in the single photon regime due to its unique capabilities in tailoring the dimensionality of the Hilbert space~\cite{Mair2001,Langford2004,Walborn2005}. In contrast, there has been very little work devoted to the OAM of intense modes described by CV; the only account in this regime is the very recent work on the creation of CV entanglement between two spatially separated OAM modes generated in atomic vapor exploiting the non-linear process of four-wave-mixing~\cite{Boyer2008science,Boyer2008}. The interest in CV OAM states stems from the further increase of the dimensionality of the Hilbert space. OAM of single photons has already found uses in various quantum information protocol, and we envisage that the CV OAM states hold similar potentials in CV quantum information but also in quantum imaging and quantum metrology. The most promising application of CV OAM states is their connectivity with atoms, thus allowing for storage of CV quantum information ~\cite{Inoue2006,Vasilyev2008}.

The simplest spatial mode that can carry OAM is the first order
Laguerre-Gaussian (LG) mode which produce either a left-handed or
right-handed corkscrew-like phase front and a ring-structured
intensity profile. They are denoted LG$_{p = 0}^{l = \pm 1}$,
where $l$ and $p$ are the azimuthal and radial mode indices.
Similarly to quantum spin angular momentum (polarization) states,
the family of first-order OAM states can be represented on a
Poincare-sphere analog which is also know as the first-order
orbital Poincare sphere~\cite{Padgett1999,Calvo2005}. The sphere
is spanned by three parameters (associated with three different
OAM states of light) that have properties identical to those of
the three polarization Stokes parameters and which we name the
orbital parameters. In this paper we demonstrate the generation of
squeezing in the orbital parameter of OAM modes using a spatially
non-degenerate optical parametric oscillator and a specially
tailored local oscillator. In addition, we measure quadrature
entanglement between the two first-order LG modes thereby
demonstrating a new type of entanglement from non-degenerate OPOs. The demonstration of similar spatially non-degenerate OPO
has recently been independently reported in Ref.~\cite{Janousek2008}.

\begin{figure}[t]
\begin{center}
\includegraphics[width=0.3\textwidth]{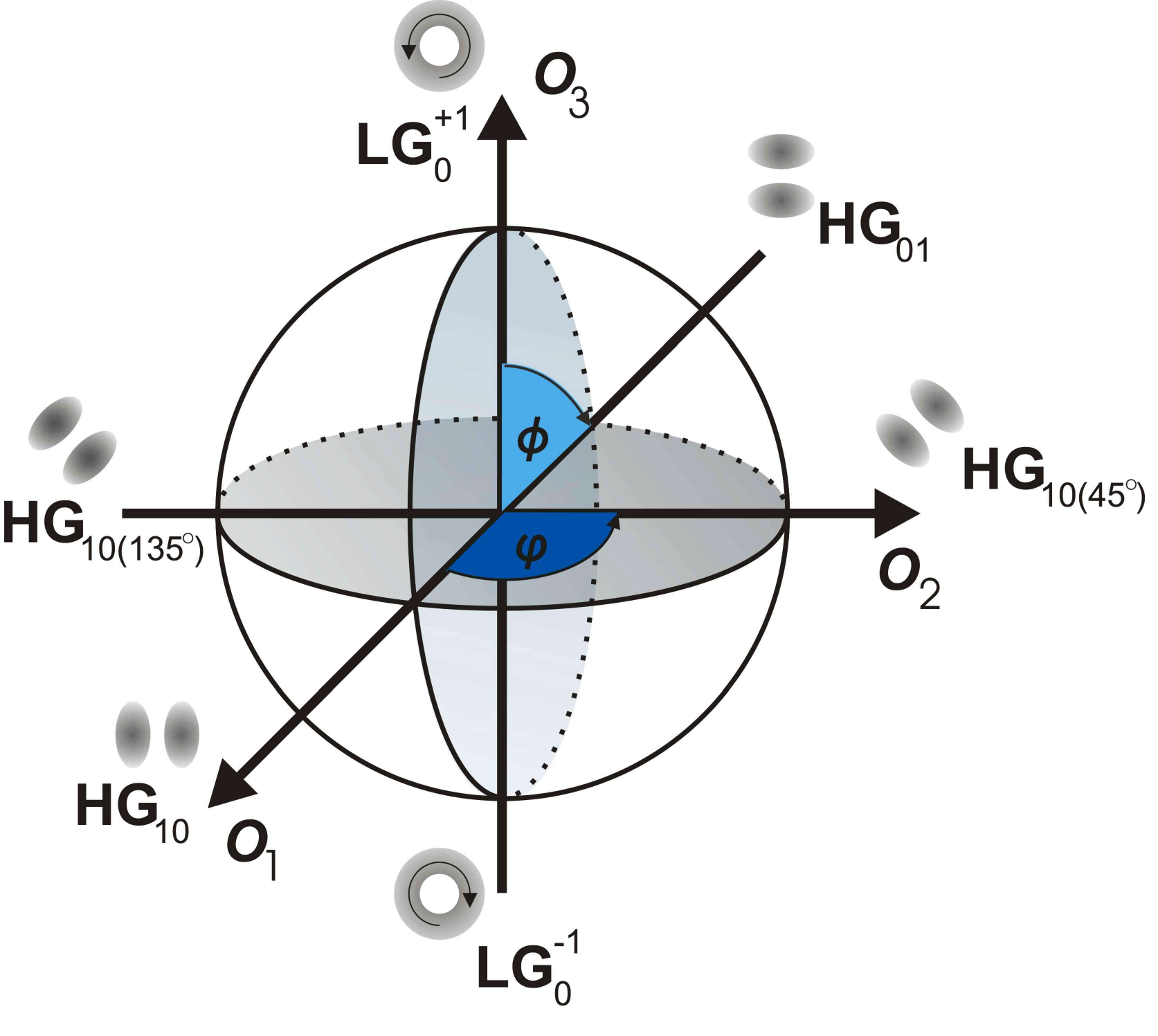}
\caption{Orbital Poincare sphere of the first-order OAM modes.
Points on the sphere are associated with a superposition of OAM
modes \cite{Calvo2005}:
$\psi_{\phi,\varphi}=\cos(\phi/2)LG_{0}^{1}+\exp(i\varphi)\sin(\phi/2)LG_{0}^{-1}$,
with azimuthal and axial angels $\phi$ and $\varphi$. }
\label{fig1poin}
\end{center}
\end{figure}

The quantum polarization degree of freedom (or spin angular
momentum) has been extensively explored and characterized in the
Schwinger representation in terms of the quantum Stokes operators
both in the two-dimensional and the infinite-dimensional Hilbert
space, where the Stokes parameter eigenvalues are either discrete
or continuous \cite{Chirkin1993,Korolkova2002,Bowen2002}. The
Stokes operators are decomposed into field operators for
orthogonal polarization modes and completely represent the quantum
dynamics of the polarization of light. Likewise, we define the
Stokes operator analogs - the orbital operators - for the first
order OAM modes as
\begin{eqnarray}
\hat{O}_1 &=& \hat{A}_{HG_{10}}^\dag \hat{A}_{HG_{10}} - \hat{A}_{HG_{01}}^\dag \hat{A}_{HG_{01}} \nonumber\\
\hat{O}_2 &=& \hat{A}_{HG_{10(45^o)}}^\dag \hat{A}_{HG_{10(45^o)}} - \hat{A}_{HG_{10(135^o)}}^\dag \hat{A}_{HG_{10(135^o)}} \nonumber\\
\hat{O}_3 &=& \hat{A}_{LG_{0}^1}^\dag \hat{A}_{LG_{0}^1} - \hat{A}_{LG_{0}^{-1}}^\dag \hat{A}_{LG_{0}^{-1}},
\label{OAMoperators}
\end{eqnarray}
where $\hat{A}^\dag$ and $\hat{A}$ are the creation and
annihilation operators for the various spatial first-order modes
given by the indices, and illustrated on the sphere in
Fig.~\ref{fig1poin}. As clearly seen from these definitions,
$\hat{O}_1$, $\hat{O}_2$ and $\hat{O}_3$  represents the
difference in the photon number between the two modes HG$_{10}$
and HG$_{01}$, HG$_{10(45^o)}$ and HG$_{10(135^o)}$, and
LG$_0^{+1}$ and LG$_0^{-1}$, respectively. Similarly to the
polarization Stokes operators, these orbital operators completely
represent the dynamics of the first-order spatial states, and thus
these operators follow the same algebra as the Stokes operators,
namely the SU(2) algebra \cite{Agarwal1999}. Likewise, the
commutation relations are $[\hat{O}_k,\hat{O}_l]=i\hat{O}_m$ where
$k,l,m\in\{1,2,3\}$ of cyclic permutation.

Spatially multimode non-classical CV states have previously been
generated in pulsed optical parametric amplification~\cite{Choi1999}, in atomic vapor~\cite{Boyer2008},
in a confocal optical parametric oscillator~\cite{Martinelli2003}
and in linear interference between different single modes
\cite{Treps2002,Lassen2007,Wagner2008}. Here we use another approach that previously has been used for the generation of single-spatial-mode squeezed states, namely a mode-stable (and non-confocal) optical parametric oscillator~\cite{Wu1986}. By employing a type I phase-matched nonlinear crystal in a mode-stable optical cavity, the polarization, the frequency and the spatial degree of freedom are usually degenerate which lead to
single mode quadrature squeezing as demonstrated by various groups. In all these experiments, the cavity supported only the
Gaussian zero'th order LG mode. However, by changing the cavity resonance frequency it is possible to generate the two first order
modes simultaneously due to their frequency degeneracy (stemming from their identical Gouy phase shifts). This means that the
down-converted signal and idler photons are produced in two distinct orthogonal spatial modes (the OAM modes, LG$_0^{+1}$ and
LG$_0^{-1}$), and thus creates quadrature entanglement between these two modes similarly to the production of entanglement
between polarization modes~\cite{Ou1992} or between frequency modes~\cite{Villar2005} in polarization or frequency
non-degenerate OPOs. The non-degeneracy of the first-order LG modes therefore adds a new member to the family of non-degenerate
OPOs capable of producing entanglement. This was also discussed in ref.~\cite{Benlloch2008} for an OPO above the oscillation
threshold. In the following, we experimentally demonstrate the generation of quadrature entangled LG modes and show that this can
be used to produce squeezing in the first order orbital parameters. Note that a thorough analysis of the transfer of OAM from the pump
to the down converted fields was carried out in Ref.~\cite{Martinelli2004}.

\begin{figure}[th]
\begin{center}
\includegraphics[width=0.4\textwidth]{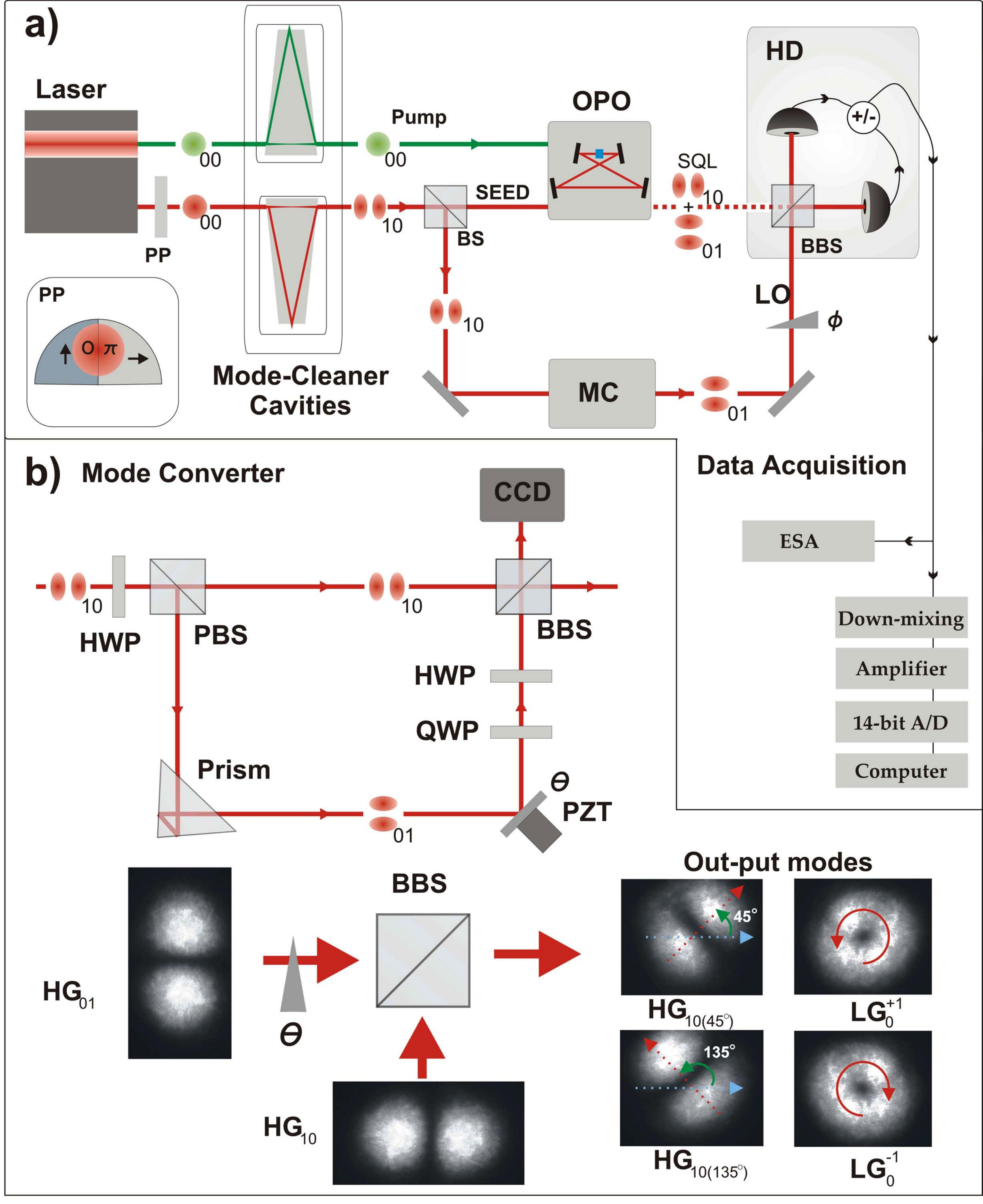}
\caption{a) Schematics of the experimental setup to generate
amplitude squeezing. b) Mode converter (MC) for the generation of
the local oscillator. The interferences process in the
MC is illustrated and some of the experimentally
obtained interference patterns are shown. BBS: Balanced beam
splitter (50/50 BS). PBS: Polarizing beam splitter. HWP:
($\lambda/2$) Half-wave plate. QWP: ($\lambda/4$) Quarter-wave
plate. PZT: Piezo-electrical element for controlling phases. PP:
Phase-Plate. The prism rotates the HG modes by 45$^o$.}
\label{fig2setup}
\end{center}
\end{figure}

Our experimental setup is depicted in Fig.~\ref{fig2setup} and
besides the laser source it consists of two mode-cleaning cavities
(for green and for infrared light (IR)), an OPO, a HG-LG mode
converter and a homodyne detection scheme. The laser source
(Diabolo from Innolight) delivers 400 mW of IR (1064 nm) light and
650 mW of green (532 nm) light. The OPO is composed of a bow-tie
shaped cavity in which a 1x2x10 mm$^3$ type I periodically poled
KTP crystal (Raicol Inc) is placed in the smallest beam waist. Our
cavity consists of two curved mirrors of 25 mm radius of curvature
and two plane mirrors. Three of the mirrors are highly reflective
at 1064~nm, $R>99.95$\%, while the output coupler has a
transmittance of $T=8\%$. At the wavelength of the pump beam (532
nm), the transmittance of the mirrors is larger than 95\%. Besides
the pump beam, the OPO cavity is seeded with a very dim HG$_{10}$
beam at 1064 nm. The spatial profile of this beam has been
tailored in the mode-cleaning cavity and to enhance the
transmission through the cavity, the Gaussian beam from the laser
passes first through a phase-flip plate (PP) which produces a
relative phase flip of $\pi$ between the two halves of the
Gaussian beam (and thus mimics the HG$_{10}$
mode)~\cite{Treps2002}. Seeding the cavity with a HG$_{10}$ mode
has a two-fold purpose; one is to enable an active cavity lock at
the frequency of the HG$_{10}$ (which coincide with the
frequencies of the HG$_{01}$, LG$_0^{+1}$ and LG$_0^{-1}$ modes),
and the other one is to ensure the generation of squeezing of the
HG$_{10}$ with a small coherent excitation (less than 1 mW).
Although the ideal spatial profile of the pump beam for down
conversion efficiency optimization is a superposition of
HG$_{00}+$HG$_{20/02}$ mode for the
HG$_{10/01}$~\cite{Lassen2007}, we have chosen to use a Gaussian
mode for simplicity reasons, and the resulting decrease in
efficiency is overcome by using a more intense beam. The relative
phase between the pump and seed is locked to de-amplification of
the seed beam, thus generating amplitude squeezing.

\begin{figure}[t]
\begin{center}
\includegraphics[width=0.4\textwidth]{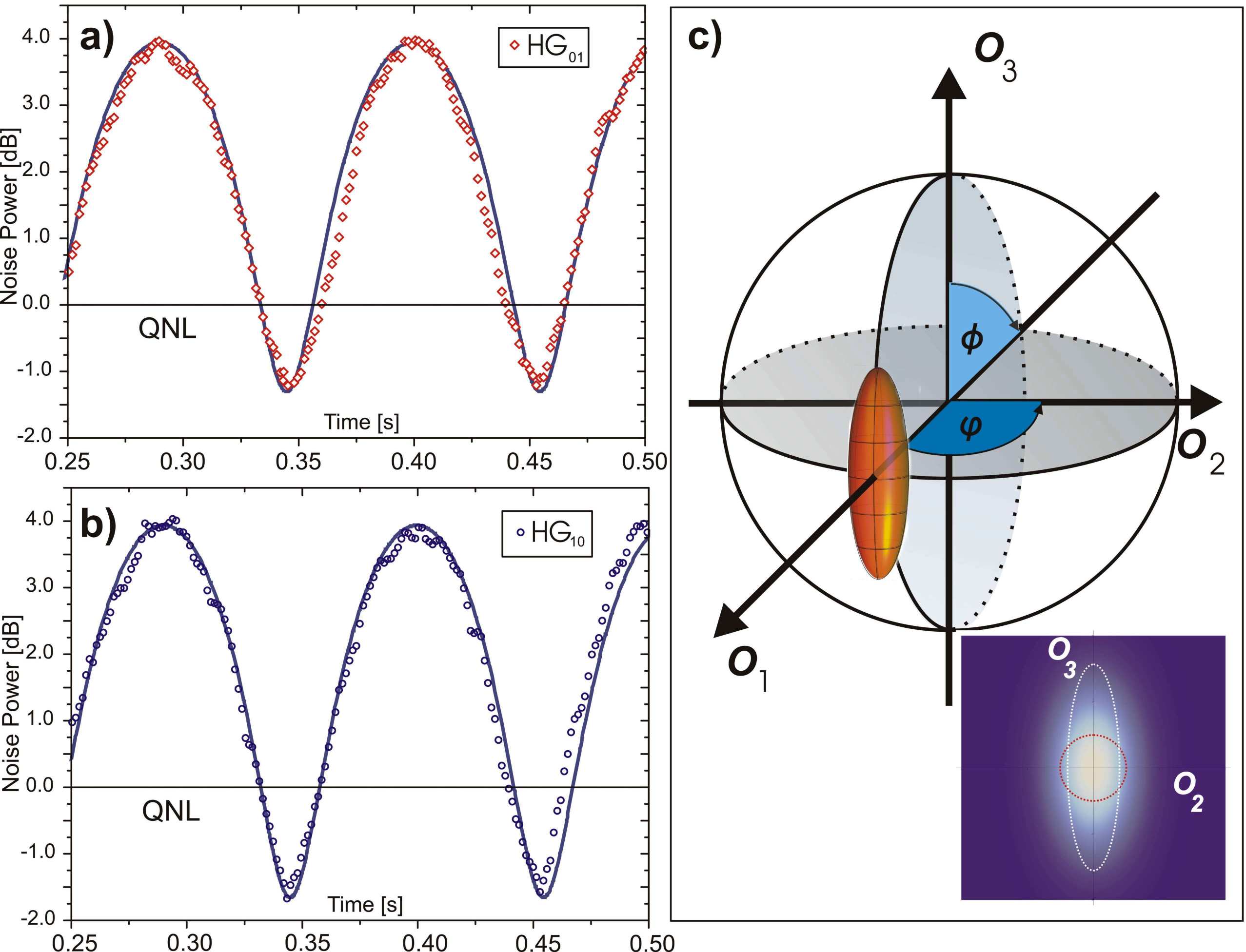}
\caption{Experimental squeezing traces for the a) HG$_{01}$ and b)
HG$_{10}$ modes, where the relative phase between the LO and the
squeezed beam is scanned. We measured $-1.6\pm0.2$~dB of
squeezing and $+4.0\pm0.2$~dB of anti-squeezing for the
HG{$_{10}$} mode, $-1.4\pm0.2$~dB of squeezing and $+4.1\pm0.2$~dB
of anti-squeezing for the HG{$_{01}$} mode. c) Illustration of the uncertainty volume on the orbital Poincare sphere and associated with the orbital parameters. The relative elongation of the volume is based on the measured values (however its size relative to the sphere is not in scale). The projection on the $O_2$-$O_3$ plane is also shown.} \label{fig3sql}
\end{center}
\end{figure}

To prove the existence of quadrature entanglement between the two
LG modes, we measure the quadrature quantum noise of the spatial
modes in a rotated basis composed of the first order HG modes,
HG$_{10}$ and HG$_{01}$: By performing a simple basis
transformation from the LG modes to HG modes, it is easy to show
that
$\hat{X}_{HG_{10}}=(\hat{X}_{LG_0^{-1}}+\hat{X}_{LG_0^{+1}})/\sqrt{2}$
and
$\hat{X}_{HG_{01}}=(\hat{P}_{LG_0^{-1}}-\hat{P}_{LG_0^{+1}})/\sqrt{2}$,
where $\hat{X}$ and $\hat{P}$ are the amplitude and phase
quadratures of the modes denoted by the lower indices. According
to the criterion of Duan et al.~\cite{Duan2000} and
Simon~\cite{Simon2000}, CV entanglement can be witnessed if
\begin{equation}
V(\hat{X}_{LG_0^{+1}}+\hat{X}_{LG_0^{-1}})+V(\hat{P}_{LG_0^{+1}}-\hat{P}_{LG_0^{-1}})<2,
\end{equation}
where $V(...)$ is the variance. Using the transformation, the criterion reduces to $V(\hat{X}_{HG_{10}})+V(\hat{X}_{HG_{01}})<2$, and thus by measuring the amplitude quadrature variances of the two HG modes, entanglement between the OAM modes can be witnessed.

The quadrature variances of the HG modes are analyzed using balanced homodyne detection (HD) with a spatially tailored Local Oscillator (LO) mode, which is either a HG$_{10}$ or a HG$_{01}$ mode depending on which signal mode is measured. We produce the HG$_{10}$ LO using the mode cleaning cavity as described above, and the HG$_{01}$ LO is generated by converting the HG$_{10}$ mode using a prism. The outcomes of the HD is fed into a spectrum analyzer which is set to display the noise power (corresponding to the second moment) of the signal beam at 5.5 MHz with a bandwidth of 300kHz (and 300 Hz averaging). We first calibrate the quantum noise limit (QNL), the result of which is illustrated in Fig.~\ref{fig3sql} by the horizontal lines. Next we measure the noise traces for the HG$_{10}$ and HG$_{01}$ modes while the phase of the LOs are scanned, see Fig.~\ref{fig3sql}. All data are normalized to the QNL.
By fitting the measured data to a theoretical squeezing curve, we find amplitude quadrature squeezing of $-1.6\pm0.2$~dB and $-1.4\pm0.2$~dB for the HG$_{10}$ and HG$_{01}$ mode, respectively (corresponding to the minima of the curves). Inserting these values into the entanglement criterion we find $V(X_{HG_{10}})+V(X_{HG_{01}})=1.42 \pm 0.01 <2$.

The measured squeezing values are degraded by the various inefficiencies of our setup.
We estimate this efficiency to be $\eta_{total}=\eta_{cav}\eta_{prop}\eta_{det}\eta_{hd}$, where
$\eta_{prop}=0.97\pm0.02$ is the measured propagation efficiency, $\eta_{det}=0.90\pm0.05$ is the measured photodiode (Epitaxx ETX500)
efficiency, {$\eta_{cav}$} = 0.94 is the estimated cavity escape efficiency and $\eta_{hd}=0.96\pm 0.02$ is the measured spatial overlap efficiency in the homodyne detector.
The total estimated detection efficiencies for our experiment is therefore $\eta_{total}=0.79\pm0.04$. From
these efficiencies can we infer the following squeezing values: -2.2$\pm0.2$ dB and -1.9$\pm0.2$ dB
for the HG$_{10}$ and HG$_{01}$ modes, respectively, and the entanglement criterion is $V(X_{HG_{10}})+V(X_{HG_{01}})=1.25 \pm0.01 <2$. We thus have proven that the spatially non-degenerate OPO produces quadrature entanglement between the first order OAM modes.

\begin{figure}[t]
\begin{center}
\includegraphics[width=0.4\textwidth]{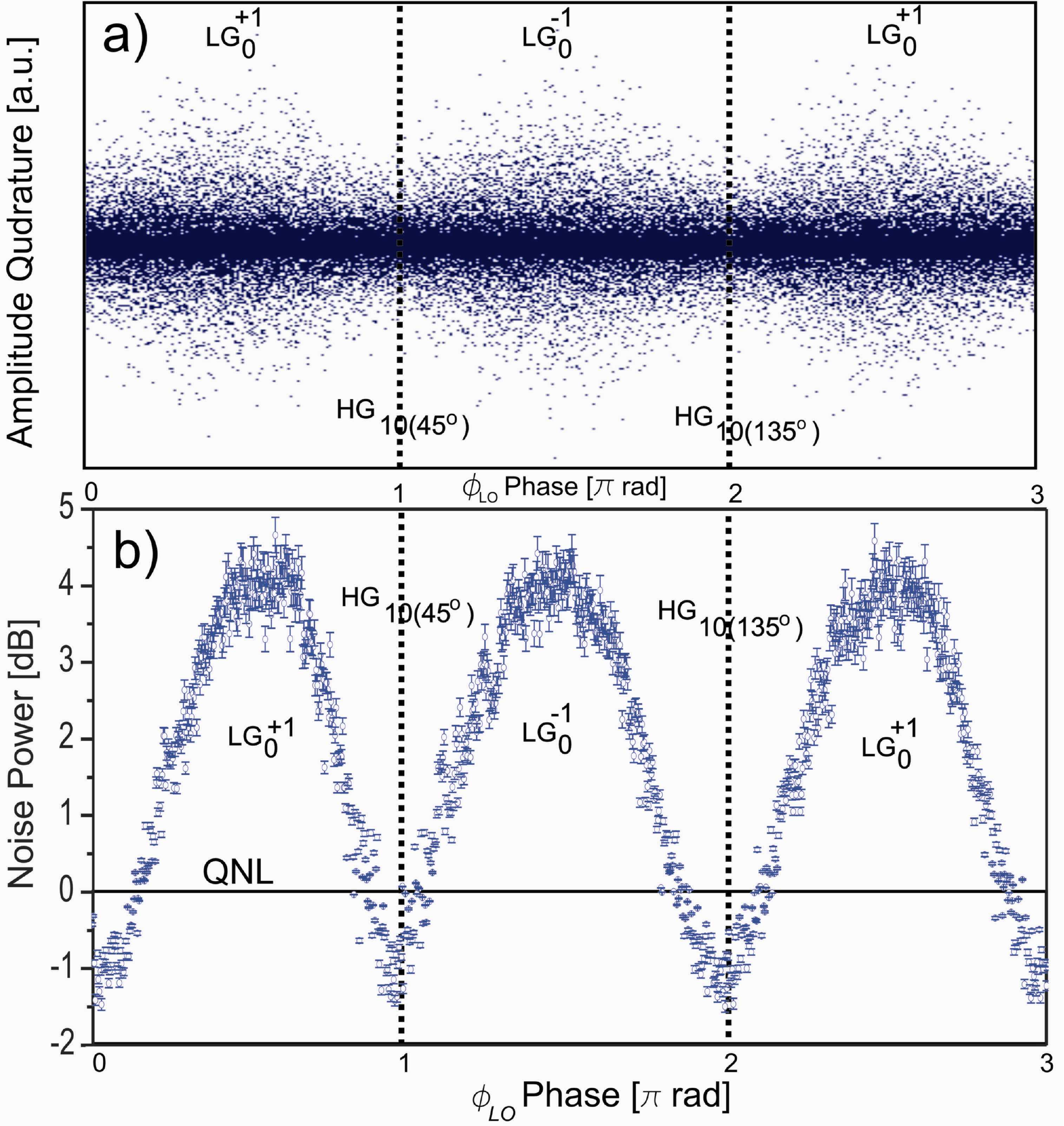}
\caption{a) Time trace for the amplitude quadrature for spatial modes along a ring on the orbital Poincare sphere spanned by $\hat{O_2}$ and $\hat{O_3}$. b) Amplitude quadrature variances for the various spatial modes on the ring. Power measurements of the mean values have been carried out simultaneously to establish the phase reference.} \label{fig4}
\end{center}
\end{figure}

We now proceed by characterizing the variances of the orbital parameters that defines the position of the state on the orbital Poincare sphere. First we note that the orbital parameters can be linearized for our setup by decomposing the field operators into a part representing the coherent excitation and a part representing the quantum noise; $\hat{A}_{HG_{01}}=\langle \hat{A}_{HG_{01}}\rangle+\hat{\delta A}_{HG_{01}}$ and $\hat{A}_{HG_{10}}=\langle \hat{A}_{HG_{10}}\rangle+\hat{\delta A}_{HG_{10}}$, and noting that $\langle \hat{A}_{HG_{01}}\rangle=0$ and $\langle \hat{A}_{HG_{10}}\rangle\neq 0$:
\begin{eqnarray}
\hat{\delta O}_1=\langle \hat{A}_{HG_{HG_{10}}}\rangle^2~\delta \hat{X}_{HG_{10}}\nonumber\\
\hat{\delta O}_2=\langle \hat{A}_{HG_{HG_{10}}}\rangle^2~\delta \hat{X}_{HG_{01}}\nonumber\\
\hat{\delta O}_3=\langle \hat{A}_{HG_{HG_{10}}}\rangle^2~\delta \hat{P}_{HG_{01}},
\label{delta}
\end{eqnarray}
where we have used the definitions $\delta \hat{X}=\hat{\delta A}+\hat{\delta A}^\dagger$ and $\hat{\delta P}=-i(\hat{\delta A}-\hat{\delta A}^\dagger)$. A similar simplification of the quantum OAM states addressing other regimes has been independently formulated in Ref.~\cite{Hsu2009}. We see that in the regime where linearization is valid, the orbital operators are linear functions of the amplitude and phase quadratures of different spatial modes. Therefore, the precision in determining first-order spatial modes depends on the quadrature noise of the light modes. For example, for a bright excitation in the HG$_{10}$ mode, as in our case, its determination on the sphere is given by the noise in the orthogonal orbital plane spanned by $\hat{O}_2$ and $\hat{O}_3$ corresponding to the quadrature noise of the HG$_{01}$ mode (see eq.~\ref{delta}). The quadrature variances of this mode is given in Fig.~\ref{fig3sql}a. Moreover the variance of $\hat{O}_1$ is given by the variance of the amplitude quadrature of the HG$_{10}$ mode given by the minimum squeezing value in Fig.~\ref{fig3sql}b. Based on these measurements we may define a cigar-shaped uncertainty volume on the orbital Poincare sphere associated with first-order OAM state as illustrated in Fig.~\ref{fig3sql}c.

Full tomographic reconstruction of the first-order OAM state is also possible by measuring all possible projections on the sphere using spatially tailored local oscillator modes.
Some modes of the LO are made by interfering equally intense HG$_{10}$ and HG$_{01}$ modes with a continuous varying relative phase shift, as shown in Fig.~\ref{fig2setup}b. By matching these LO modes with the OAM modes, we measure the amplitude quadratures of the modes along a ring on the orbital sphere, thus mapping out the quadrature noise of a whole family of different OAM states. The results of the amplitude quadrature measurements are shown in Fig.~\ref{fig4}a for a 150 kHz broad signal at 5.5 MHz. We also compute the variance of these data as depicted in Fig.~\ref{fig4}b.

To conclude, we have generated a new quantum state of light composed of quadrature entangled LG modes. For the generation we used an OPO operating in a new regime where all field parameters are degenerate except for its spatial degree of freedom for which it is two-fold degenerate. The produced OAM states from the OPO are mapped on an orbital Poincare sphere which is similar to the standard Poincare sphere for polarization. We generate a classically bright HG$_{10}$ mode described on this sphere but its exact location is determined by the noise of the orthogonal HG$_{01}$ dark mode. As this dark state was quadrature squeezed from the OPO, we have demonstrated squeezing in the orbital parameters defining the position of the mode on the orbital Poincare sphere. In addition we have measured the noise of the bright orbital parameter and thereby reconstructed the uncertainty volume of the first-order OAM state.

We acknowledge the financial support from the EU (COMPAS) and the Danish Research Council (FTP). ML acknowledge support from the Alexander von Humboldt Foundation.

\end{document}